\documentclass[conference]{IEEEtran}

\usepackage{url}
\usepackage{cite}
\usepackage{graphicx}
\usepackage{amsmath}
\usepackage{subfigure}
\usepackage{color}
\usepackage{comment}
\usepackage[ruled,figure,linesnumbered]{algorithm2e}

\begin{document}

\title{Deep and Shallow convections in Atmosphere Models on Intel{\textregistered} Xeon Phi{\texttrademark} Coprocessor Systems}
\author{\IEEEauthorblockN{$^1$Srinivasan Ramesh, $^2$Sathish Vadhiyar, $^{3,4}$Ravi Nanjundiah, $^4$PN Vinayachandran}
\IEEEauthorblockA{
$^1$Department of Computer and Information Science, University of Oregon, USA \\
$^2$Department of Computational and Data Sciences, Indian Institute of Science, Bangalore, India \\
$^3$Indian Institute of Tropical Meteorology, Pune, India \\
$^4$Centre for Atmospheric and Oceanic Sciences, Indian Institute of Science, Bangalore, India \\
sramesh@cs.uoregon.edu, vss@cds.iisc.ac.in, ravisn@tropmet.res.in, vinay@caos.iisc.ernet.in
}
}

\maketitle

\begin{abstract}
Deep and shallow convection calculations occupy significant times in atmosphere models. These calculations also present significant load imbalances due to varying cloud covers over different regions of the grid. In this work, we accelerate these calculations on Intel{\textregistered} Xeon Phi{\texttrademark} Coprocessor Systems. By employing dynamic scheduling in OpenMP, we demonstrate large reductions in load imbalance and about 10\% increase in speedups. By careful categorization of data as private, firstprivate and shared, we minimize data copying overheads for the coprocessors. We identify regions of false sharing among threads and eliminate them by loop rearrangements. We also employ proportional partitioning of independent column computations across both the CPU and coprocessor cores based on the performance ratio of the computations on the heterogeneous resources. These techniques along with various vectorization strategies resulted in about 30\% improvement in convection calculations.
\end{abstract}


\section{Introduction}
\label{intro}

The importance of climate to energy usage and agriculture has made it a prominent field of study. Climate study is based on the mathematical models of physical processes such as radiation, circulation, precipitation and their interaction with chemical and biological processes. The equations of conservation of mass, momentum, energy and species are used to represent various components of the climate model such as atmosphere, land, sea and ice. Numerical methods are extensively used to solve these equations and as a result, climate models have many computationally intensive routines.

One  such  climate  model is the Community Earth System Model \cite{hurrell-cesm-bams2013}, developed and maintained by the National Center for Atmospheric Research (NCAR). CESM consists of several component models, eg. physical climate, chemistry, land ice, whole atmosphere, etc, that can be coupled in different configurations. In all cases, geophysical fluxes across the components are exchanged via a central coupler module. A large number of simulations with CESM have been conducted, some of which are available for community analysis \cite{hurrell-cesm-bams2013}.

\begin {figure}
\centering
\includegraphics[scale=0.2]{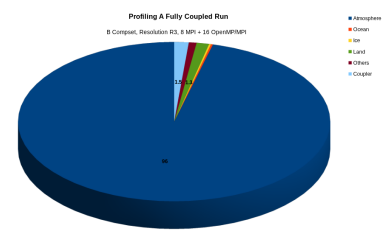}
\caption{Execution Profile of CESM}
\label{cesm_profile}
\end{figure}

The atmosphere is the most time-consuming model in CESM, as shown in Figure \ref{cesm_profile}.  We obtained such execution profiles using Intel VTune Amplifier and HPCToolkit profiler\cite{hpctoolkit}. The execution profile is obtained by running CESM v 1.2.2 with $f02\_g16$, the highest resolution\footnote{see Table \ref{resolutions} for resolutions.} and B compset (fully-coupled run) on eight-node 128-core Intel Xeon processors with 8 MPI processes and 16 OpenMP threads per process for a total of 128 threads. The model used for atmosphere is the Community Atmosphere Model (CAM5)\cite{collins-cam-jc2006,neale-meanclimate-jc2013}. CAM consists of two computational phases, namely, dynamics and physics. The dynamics advances the evolutionary equations for the flow of atmosphere and the physics approximates sub-grid phenomena including clouds, long and short wave radiations, precipitation processes and turbulent mixing. The default core for the dynamics is a finite volume method \cite{neale-meanclimate-jc2013} that uses a longitude $\times$  latitude $\times$ vertical level computational grid over the sphere. The physics in CAM is based upon vertical columns whose computations are independent from each other. The parallel implementation of the physics is based on the assignment of columns to MPI processes and using OpenMP threads within a process to compute the columns assigned to a process.

\begin {figure}
\centering
\includegraphics[scale=0.2]{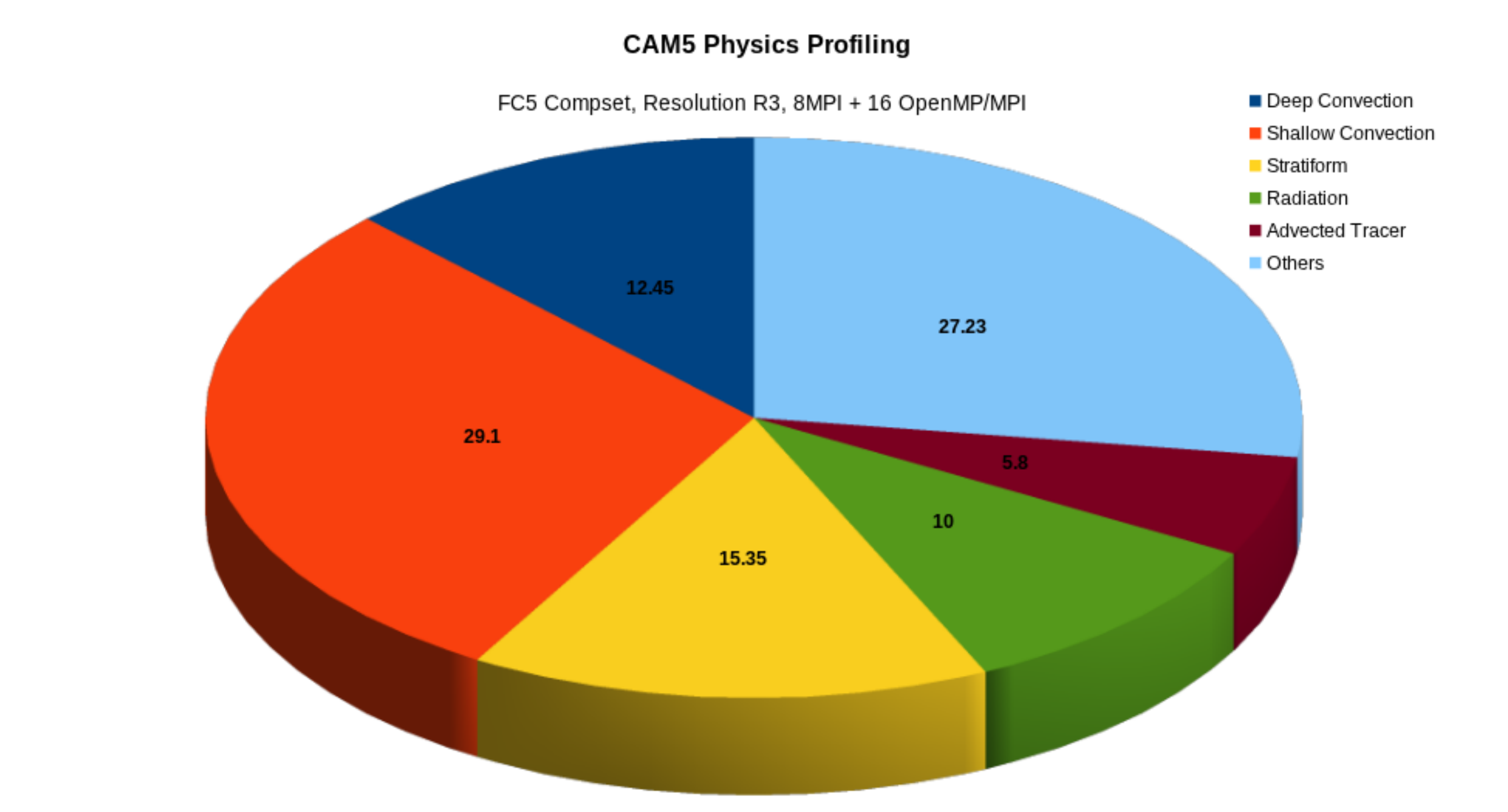}
\caption{Execution Profile of the CAM5 Physics model}
\label{phy_profile}
\end{figure}

Most of the work to accelerate climate science routines have concentrated on the dynamics. In this work, we chose to accelerate the physics routines as they were observed to consume significant times, typically about 40\% of the entire CESM run. Figure~\ref{phy_profile} illustrates the execution profile of CAM5 physics. This figure shows that the the deep and shallow convection routines (dark blue and orange) are the most time consuming routines, taking about 41\% of the total time spent in the CAM5 physics. Deep convections consider parameterization of clouds on the basis of moist static instability over the entire atmosphere column. Shallow convections consider clouds when the moist static instability is present only over parts of the atmosphere column.  These calculations also present significant load imbalances due to varying cloud covers over different regions of the grid. Convection is at the crux of modeling clouds which play an important role in atmospheric circulation. More complex methods such as super parameterization, while being more accurate could be about two orders of magnitude more time-consuming than present day approaches. Thus, acceleration of convection calculations can result in use of more sophisticated methods.

Accelerators and co-processors are widely prevalent and have been used to provide high performance for many scientific applications. Intel{\textregistered} Xeon Phi{\texttrademark} coprocessors have been gaining ground to provide speedups for advanced scientific applications \cite{liu-swaphils-cluster2014,heybrock-latticeqcd-sc2014,luo-mica-bmc2015}. However, the use and demonstration of these coprocessors for climate modeling are limited. In this work, we accelerate the deep and shallow convection calculations on Intel{\textregistered} Xeon Phi{\texttrademark} Coprocessor Systems. By employing dynamic scheduling in OpenMP, we demonstrate large reductions in load imbalance and about 10\% increase in speedups. By careful categorization of data as private, firstprivate and shared, we minimize data copying overheads for the coprocessors. We identify regions of false sharing among threads and eliminate them by loop rearrangements. We also employ proportional partitioning of independent column computations across both the CPU and coprocessor cores based on the performance ratio of the computations on the heterogeneous resources. These techniques along with various vectorization strategies resulted in about 30\% improvement in convection calculations.

In Section \ref{back}, we give the overall structure of the atmosphere model and the convection calculations.  Section \ref{related} covers related work in the area of high performance of climate and weather models on accelerators and coprocessors. In Section \ref{opt}, we describe our different optimization techniques including avoiding false sharing, and proportional partitioning of the computations among the CPU and Xeon Phi cores. Section \ref{exp_res} presents experiments and results. In Section \ref{discussion}, we derive general principles from our optimizations on the climate modeling application. Section \ref{con_fut} gives conclusions and presents scope for future work.

\section{Background}
\label{back}

\subsection{Convection Calculations}

The computational grid for the atmosphere is divided into latitudes, longitudes along the x and y direction and vertical columns along the z direction. The vertical columns extend from the surface up through the atmosphere. The columns are further divided into layers. The characteristic feature of the physics routines is that every column can be computed independent of the other columns giving rise to data parallelism that can be exploited on multi and many core architectures. For the purpose of load balancing among the different CPU processes, the columns are grouped into {\em chunks}. A chunk is a unit of execution containing a configurable number of columns. A chunk may or may not contain contiguous set of vertical columns i.e. columns from neighboring grid points may not be in a single chunk. These chunks are distributed among the MPI processes and the OpenMP threads. Every physics routine is called for each chunk. The pseudo-code for physics calculations is shown below.

\begin{algorithm}
\begin{small}
\SetKwInOut{Input}{input}
\SetKwInOut{Output}{output}
\SetKwFunction{stratiform}{stratiform}
\SetKwFunction{deepconvection}{deep\_convection}
\SetKwFunction{shallowconvection}{shallow\_convection}
\SetKwFunction{radiation}{radiation}

  \ForEach{time step}{
    \ForEach{chunk}{
      \stratiform() \;
      \deepconvection() \;
      \shallowconvection() \;
      ... \BlankLine
      \radiation() \;
      ...
    }
  }
\caption{Pseudocode for Physics Calculations in CAM}
\label{physics_pseudo}
\end{small}
\end{algorithm}

Convections are of two kinds: deep and shallow convections. \\
{\bf Deep Convection:} In tropics tall clouds occur on scales of 1-10 km much smaller than that of the resolved grid (typically about 50 km in climate models).  These clouds could be within thunderstorms or similar unsettled weather conditions.  While they cannot be resolved explicitly by the model, their effect is very important and needs to be incorporated. Hence they are parameterized. The parameterization in most climate models works on the basis of moist static instability of the atmospheric column. Formation of clouds removes the moist static instability and cause precipitation (which could be rain or snow). There are various schemes to parameterize these tall clouds. In CAM5 they are parameterized using the method of Zhang and McFarlane \cite{zhang-sensitivity-atmocn1995}. \\
{\bf Shallow Convection:} If the moist static instability is not present over the entire atmospheric column but over only part of it, clouds will form only in this part. Such a situation is handled by shallow convection routines. Shallow convection may or may not cause precipitation. In CAM5, shallow convection is handled using the University of Washington scheme \cite{park-UWshallowconvect-jc2009}.

Along  with  other  processes  like radiation  and  stratiform,  deep  and  shallow  convection  are  two  physical  processes that  are  applied  on  all  grid  points  before  coupling  to  land,  sea  and  ice  models.  Computations in these routines are related to local conditions and it is difficult to a-priori predict their occurrence. It is also difficult to determine a pattern in their occurrence -- though deep convection could be more prominent in the tropics than in extra-tropics. Such computations are therefore not conducted at every grid point at a particular time-step. The stencil of these computations would change with time and hence would be difficult to develop load-balancing techniques for these computations. Hence it is essential that these routines be accelerated.  Shallow and deep convection computations are parallel across chunks and the loops  within  the  shallow  and  deep  convection  routines  that  represent  the  core computations  iterate  over  columns  within  a  chunk. The chunk computations within an MPI process are parallelized across OpenMP threads.

The  core computations in deep convection consists of two loops - entrainment and precipitation loops, as shown in Figure \ref{deep_convect}.
Entrainment is a process in which dry air from outside a cloud mixes with the moist (almost saturated) air inside the cloud. This process reduces the relative humidity of the air and could result in shorter clouds of smaller heights. Precipitation is the process of conversion of water vapor inside the cloud converting into liquid droplets that coalesces and form rain. The entrainment and precipitation loops iterate over columns in a given chunk and the computations are independent across different columns/iterations. 
In this work, we exploit the fine-level parallelism of the column computations on Xeon Phi by adopting column-level parallelism for the convection routines and the default chunk-level parallelism for the other routines.

\begin{algorithm}
\begin{small}
\SetKwInOut{Input}{input}
\SetKwInOut{Output}{output}
\SetKwFunction{ientropy}{ientropy}
\SetKwFor{Do}{do}{}{end}

  \tcc{Entrainment Loop}
  \Do{i=1,numColumnsinChunk}{
    \Do{k=1,numVerticalLevels}{
      ... \BlankLine
      ... \BlankLine
      \ientropy() \;
      ...
    }
  }
  \BlankLine

  \tcc{Precipitation Loop}
  \Do{i=1,numColumnsinChunk}{
    \Do{k=1,numVerticalLevels}{
      ... \BlankLine
      ... \BlankLine
      \ientropy() \;
      ...
    }
  }

\caption{Deep Convection}
\label{deep_convect}
\end{small}
\end{algorithm}

\subsection{Intel Xeon Phi Architecture}

Intel's first generation Many Integrated Core (MIC) architecture codenamed Knights Corner, is an x86 based system that contains up to 61 in-order processing cores, and offers peak double precision performance of nearly 1.2 TFLOPS. Each of these cores supports up to 4-way hardware multithreading and features a vector unit which uses wide 512 bit registers. Each core features fully coherent L1 and L2 caches, with a bidirectional ring that provides fast access to L2 caches of other cores.

The Xeon Phi cards used in our experiments offers a peak memory bandwidth of nearly 352 GBPS, but the L1 and L2 caches offer much higher memory bandwidth than the memory, and using them effectively is the key to approaching peak performance. Apart from thread parallelism, it is crucial to properly utilize the wide vector unit; more so than on the Xeon which has only 256 bit wide registers. Although the Xeon Phi offers gather and scatter operations for vectors, vectorization is significantly more effective if data is contiguous in memory with unit stride accesses. Data alignment will provide even better vector performance. The Xeon Phi also features low precision hardware support for certain math functions like power, logarithm etc apart from FMA (Fused Multiply and Add) that provide additional speedup for workloads that have these computations.

Programming the Xeon Phi is similar to programming any other x86 machine - the same programming model is applicable with a host of popular libraries like MPI and OpenMP that are supported. The same optimizations that apply on the Xeon also apply without change on the Xeon Phi. In general, development time of a parallel application on the Xeon Phi is short.  The Xeon Phi offers three modes of operation - native, offload and symmetric. We use the offload model of execution in our study, where the host offloads a portion of computation to the Xeon Phi either synchronously or asynchronously with simultaneous CPU executions.

\section{Related Work}
\label{related}

There have been a number of efforts in using GPUs for climate and weather models.  Michalakes and Vachharajani \cite{michalakes-gpuweather-ipdps2008} used GPUs to improve the performance of the Weather Research and Forecast (WRF) model.  Their work resulted in  in 5-20x speed-up for the computationally intensive routine WSM5. 
In the work by Govett et al. \cite{govett-nim-ccgrid2010}, the non-hydrostatic icosahedral (NIM) model was ported to the GPU \cite{govett-nim-ccgrid2010}.  The dynamics portion which is the most expensive part of the NIM model was accelerated using GPU and the speed-up achieved was about 34 times on Tesla - GTX-280 when compared to a CPU.
The Oak Ridge National Laboratory (ORNL) ported the spectral element dynamical core of CESM, HOMME, to the GPU \cite{carpenter-homme-ijhpca2013}.
A very high resolution of (1/8) th degree was used as a target problem.
Using asynchronous data transfer, the most expensive routine performed three times faster on the GPU than the CPU. This execution model was shown to be highly scalable.  The climate model ASUCA \cite{shimokawabe-asuca-sc2010} is a production weather code developed by the Japan Meteorological Agency. By porting their model fully onto the GPU they were able to achieve 15 TFlops in single precision using 528 GPUs. The TSUBAME 1.2 supercomputer in Tokyo Institute of Technology was used to run the model. The CPU is used only for initializing the models and all the computations are done on the GPU. There are different kernels for the different computational components.

In the work by Schalkwijk et al. \cite{schalkwijk-eddygpu-ams2015}, the authors have utilized the GPU's for Large Eddy Simulation (LES) models, which has allowed them to provide turbulence-resolving numerical weather forecasts over a region the size of the Netherlands, at 100m resolution. Garcia et al. \cite{garcia-cloud-iccs2012} accelerate a Cloud Resolving Model (CRM) by implementing the MPDATA algorithm on GPU using CUDA.  They  perform optimizations like data reuse on GPU for saving transfer time, coalesced memory accesses on GPU's, and utilizing the texture memory and shared memory on the GPU. Fuhrer et al. \cite{fuhrer-portableweatherclimate-sci2014} optimize the atmospheric model COSMO by rewriting the dynamical core using STELLA DSEL and porting the remaining parts of the Fortran code to the GPU's using OpenACC compiler directives.

Intel Xeon Phi is one of the important accelerator architectures and have been becoming prevalent in supercomputer sites that adopt heterogeneous systems. Hence, porting climate models to Intel Xeon Phi accelerators is essential.
Intel Xeon Phi processors have been used to provide high performance for different scientific domains \cite{liu-swaphils-cluster2014,heybrock-latticeqcd-sc2014,luo-mica-bmc2015}.
There have been recent efforts in porting weather and climate models on Intel Xeon Phi accelerators. Mielikainen et al. have a number of efforts on optimizing Weather Research Forecast Model (WRF) on Intel Xeon Phi architecture. In \cite{mielikainen-optwrf-spie2014, mielikainen-revisitingCloud-spie2015}, the authors have  optimized  the  Thomspson  cloud  microphysics scheme,  a sophisticated cloud microphysics scheme.  They have used optimization techniques such as modifying the tile size processed by each core, using SIMD, data alignment, memory footprint reduction, etc.  to achieve a speedup of 1.8x over the original code on Intel Xeon Phi 7120P.
 and 1.8x over the original code on dual socket configuration of eight core Intel Xeon E5-2670. In  another work \cite{mielikainen-longwave-spie2015}, the authors optimize the longwave radiative transfer scheme of the Goddard microphysics scheme of the WRF model, for Intel MIC architecture.  Their optimization yields a speedup of 2.2x over the original code on Xeon Phi 7120P. They also optimize the updated Goddard shortwave radiation of the WRF model for Intel Xeon Phi \cite{mielikainen-shortwave-spie2015}.  They observe a speedup of 1.3x over the original code on Xeon Phi 7120P. 

Betro et al. \cite{betro-performancemetrics-cug2013} highlight experiences and knowledge gained from porting such codes as ENZO,  H3D, GYRO, a  BGK Boltzmann solver,  HOMME-CAM, PSC, AWP-ODC, TRANSIMS, and ASCAPE to the Intel Xeon Phi architecture running on a Cray CS300- AC Cluster Supercomputer named Beacon.
Most of these were ported by compiling with the flag -mmic.  They conclude that accelerator based systems are the wave of the future based both on their power consumption and variety of programming paradigms to  fit the needs of all application developers. 

Michalakes et. al \cite{michalakes-optimizingweathermodel-hmcw2014} optimize a standalone kernel implementation of Rapid Radiative Transfer Model of the NOAA Nonhydrostatic Multiscale Model (NMM- B). They apply methods such as dynamic load balancing, lowering inner loops, avoiding vector remainders, trading computation for data movement, prefetching etc.  and obtain a speedup of 1.3x over the original code on Xeon Sandybridge and 3x over the original code on Intel Xeon Phi.

To our knowledge, ours is the first effort on accelerating convection calculations on Intel Xeon Phi clusters. While existing efforts on accelerators and co-processors focused on data management and vectorization, in addition to these optimizations, our work proposes novel asynchronous execution model for simultaneous executions on both CPU and coprocessor cores.

\section{Methodology}
\label{opt}

In our work, we explore fine-grained  parallelization  across the column iterations for both deep convection, shown in Figure \ref{deep_convect}, and shallow convection. We propose various optimizations including load balancing, avoiding false sharing, data management and proportional partitioning of the columns across both the Xeon CPU and Xeon Phi coprocessor cores, shown in Figure \ref{deep_convect}, and shallow convection. We also propose various other  optimizations including proportional partitioning of the shallow convection columns across both the Xeon CPU and Xeon Phi coprocessor cores.

\subsection{Load Balancing Convections}

Deep convection is bound to have load imbalances due to varying cloud cover among the columns owned by the OpenMP threads. At grid points where there is no deep clouding, the corresponding column iterations of the deep convection computations will be skipped. It is likely that deep convection occurs over domains computing the tropics rather than those of the extra-tropics and polar regions. While CESM supports load balancing at the chunk level for the overall physics calculations, our work explores load balancing for fine-grained balancing at the column level. We have employed OpenMP {\em dynamic scheduling} to distribute the columns/iterations to the threads in a load balanced manner. 

However, dynamic scheduling incurs overheads since the OpenMP runtime has to dynamically schedule the column iterations to the next available and free threads. The OpenMP dynamic chunksize parameter ({\em omp chunk size}) also plays a role in determining the scheduling overhead. To offset the overheads, the number of column iterations and/or the workload per column has to be sufficiently large. Thus the performance benefits of dynamic scheduling depend on the number of columns per chunk ({\em the model chunk size}) and the model resolution used. Table \ref{omp_chunk_xeon} shows the effect of OpenMP dynamic chunk sizes on the loop runtimes for the R1 resolution with 5 days of simulation on the Xeon CPU with 16 threads. The results indicate that increasing the chunk size of the Xeon beyond 14 degrades performance due to dynamic scheduling overheads. On Xeon Phi with 240 threads, using OpenMP dynamic chunk sizes of 1, 2 and 3 yielded loop runtimes of 311, 357 and 405 seconds, respectively. Due to the large number of Xeon Phi threads and limited number of columns per chunk, we were unable to increase the chunk size beyond 3. Thus, we generally found that an omp chunk size of 10-14 for the Xeon, and 1-2 for the Xeon Phi produced the best results.  We increased the model chunk size from the default value of 16 to 864 for the R1 resolution, and 1728 for the R2 and R3 resolutions for the dynamic scheduling to be effective.

\begin{table}
\small
\centering
\begin{tabular}{||p{0.58in}|p{0.15in}|p{0.15in}|p{0.15in}|p{0.15in}|p{0.15in}|p{0.15in}|p{0.15in}|p{0.15in}|p{0.15in}|p{0.15in}||}
\hline\hline
{\it Chunk Size} & 1 & 2 & 4 & 8 & 10 & 12 & 14 & 16 & 20 \\
\hline
{\it Runtime (secs)} & 117.8 & 104.3 & 92.4 & 73.8 & 73.1 & 74.7 & 72.0 & 76.2 & 76.7 \\
\hline\hline
\end{tabular}
\caption{Effect of OpenMP dynamic chunk sizes on loop runtime on Xeon}
\label{omp_chunk_xeon}
\end{table}

Our single-node experiments using 16 threads on Xeon and 240 threads on Xeon Phi show that the use of OpenMP dynamic scheduling resulted in improvements of 11.8\% on Xeon and 17.6-20.1\% on Xeon Phi for both the entrainment and precipitation loops when compared to the default static scheduling of the threads.

Shallow convection too displays significant load imbalance among the column computations. In areas where there are no clouds, the cumulus convection computations are skipped, and this leads to load imbalance. Based on certain thermodynamic properties including pressure, a condition variable is assigned to either true or false for each column. This condition is set at multiple places in the loop, and if at any time it is true, further computations for that column are stopped. This leads to differing amounts of computations in different columns based on the condition, resulting in load imbalance.
This is depicted in Figure \ref{shallow_convect}.
Similar to deep convection, we employed OpenMP dynamic scheduling for the shallow convection and improved the performance of the loop by 15\% on the Xeon, and 10.5-12.5\% on the Xeon Phi in our single-node experiments.

\begin{algorithm}
\begin{small}
\SetKwInOut{Input}{input}
\SetKwInOut{Output}{output}
\SetKwFor{Do}{do}{}{end}

  OMP PARALLEL SCHEDULE(DYNAMIC,2)
  \Do{i=1,numColumnsinChunk}{
      ... \BlankLine
      ... \BlankLine
      shouldColumnExit = checkThermoDynProperties() \;
      \If{shouldColumnExit = TRUE}{
        goto END \;
      }
      \tcc{The above condition check is repeated at multiple points in the routine. Some column computations exit early leading to load imbalance}
      ... \BlankLine
      ... \BlankLine
      shouldColumnExit = checkThermoDynProperties() \;
      \If{shouldColumnExit = TRUE}{
        goto END \;
      }
      ... \BlankLine
      \tcc{Write output variables for this column}
      END: \BlankLine
        \tcc{Set output variables to ZERO for this column}
    }
\caption{Shallow Convection}
\label{shallow_convect}
\end{small}
\end{algorithm}

\subsection{Data Management for Offloading}

After solving the load imbalance problem in the Shallow convection loop, we found a large difference of about 7-25\% between the time taken for the entire OpenMP column-loop and the individual thread times that correspond to the useful work done by the threads. Further analysis using Intel Vtune suggested that this was due to memcpy operation as a result of using the firstprivate clause for a large number of local variables in the OpenMP loop. The shallow convection routine had about 500 such local variables and about 40 input/output array variables. Out of the 500 local variables, about 220 are array variables. Total size of all the local 500 variables including scalars and arrays are roughly 16 MB, out of which the arrays contributed 99\%. FirstPrivate for a variable in OpenMP results in private copies of the variable for each thread and initialized to the value of the variable in the master thread prior to the OpenMP parallel loop. This incurs large overheads in OpenMP runtime since the runtime performs memcpy operations to initialize the variable copies in the threads. This memcpy time grew with the number of threads used, and severely affected Xeon Phi performance due to larger number of threads (180) used on Xeon Phi. The memcpy time alone contributed to about 44\% of the runtime for the loop on the Xeon Phi.

All the 500 local variables could not be designated with the default {\em private} setting of OpenMP, since about 100 of the variables were initialized before the loop began. Using shared for all the variables also yielded incorrect results. Upon closer inspection, we found that all except 4 of the 100 variables that were being initialized were diagnostic array variables that were not used in model output. As each thread was indexing into the arrays, these variables were designated as shared. Only the 4 scalar variables were truly firstprivate since they were being initialized and used inside the loop. Thus, most of the large arrays could be safely used as shared variables, and many scalars could be used as private variables (instead of firstprivate) in the OpenMP clause for the shallow convection loop. This resulted in drastic reduction of the number of variables declared as firstprivate to only the four scalar variables.

Hence, we proceeded to use dynamic scheduling for the shallow convection loop. We also verified the correctness by checking bit-by-bit accuracy. We also explored the use of OpenMP's {\em task} construct in place of the for-loop parallelization. We found that this resulted in a performance slowdown of 1.2X when compared to dynamic scheduling on both Xeon and Xeon Phi.

\subsection{False Sharing among Threads and Loop Rearrangement}

Further studying the scalability bottlenecks and hotspots with Intel VTune, we found that the shallow convection routine has two loops whose scaling efficiency and speedups are only about 37\% and 6x, respectively, when executing with 16 threads on the Xeon CPU cores. The two loops also contributed to about 35\% of the overall shallow convection routine. These loops write output values to 2D-arrays of $cols\times verticals$ for various vertical levels corresponding to the columns possessed by a thread as shown in Figure \ref{falseshare_write}. Due to the column major traversal in Fortran, multiple threads owning different columns access successive memory locations corresponding to the same vertical level, resulting in false sharing of cache lines by the threads. Also, access of successive vertical levels for a column by a thread results in access with non-unit large strides. As confirmed by the Intel compiler optimization report, this results in inefficient vectorization of the loop with generation of scatter-gather instructions.

\begin{algorithm}
\begin{small}
\SetKwInOut{Input}{input}
\SetKwInOut{Output}{output}
\SetKwFor{Do}{do}{}{end}

  \Output{A(numColsinChunk,numVerticalLevels), B(numColsinChunk,numVerticalLevels)}

  OMP PARALLEL SCHEDULE(DYNAMIC,2)
  \Do{i=1,numColumnsinChunk}{
      ... \BlankLine
      ... \BlankLine
      \tcc{Write output variables for this column}
      \Do{k=1,numVerticalLevels}{
        A(i,k) = ....\;
        B(i,k) = ....\;
      }
      END:
      \tcc{Set output variables to ZERO for this column}
      \Do{k=1,numVerticalLevels}{
        A(i,k) = 0 \;
        B(i,k) = 0 \;
      }
    }
\caption{False Sharing and Unvectorized Writing Loops}
\label{falseshare_write}
\end{small}
\end{algorithm}

The solution to both the false sharing problem and inefficient vectorization is to interchange the array dimensions, and pad the arrays such that no two threads write to the same cache line at any time. Specifically, the number of vertical levels (k-loop in Figure \ref{falseshare_write}) is 30 in CESM. Since the cache line in Xeon and Xeon Phi architectures is 64 bytes, we need to ensure that the size of the innermost dimension, corresponding to the vertical levels, in arrays $A$ and $B$ is a multiple of 64. Hence, padding of two elements for the innermost dimension was applied. The modified code is shown in Figure \ref{nofalseshare_write}.

\begin{algorithm}
\begin{small}
\SetKwInOut{Input}{input}
\SetKwInOut{Output}{output}
\SetKwFor{Do}{do}{}{end}

  \tcc{Interchange of array dimensions}
  \Output{A(numVerticalLevels+padding,numColsinChunk), B(numVerticalLevels+padding,numColsinChunk)}

  OMP PARALLEL SCHEDULE(DYNAMIC,2)
  \Do{i=1,numColumnsinChunk}{
      ... \BlankLine
      ... \BlankLine
      \tcc{Write output variables for this column}
      \Do{k=1,numVerticalLevels+padding}{
        A(k,i) = ....\;
        B(k,i) = ....\;
      }
    END:
      \tcc{Set output variables to ZERO for this column}
      \Do{k=1,numVerticalLevels+padding}{
        A(k,i) = 0 \;
        B(k,i) = 0 \;
      }
    }
\caption{Modified code to avoid false Sharing and improve vectorization}
\label{nofalseshare_write}
\end{small}
\end{algorithm}

Our optimization resulted in the single-thread performance improvement, as mentioned in Section \ref{exp_res}, due to increased vectorization. This was also confirmed by the compiler report that did not show generation of scatter-gather instructions. 

\subsection{Proportional Partitioning of Column Computations across CPU and Co-processor Cores}

Recalling that the column computations are independent across different columns, all the columns corresponding to a MPI task of a node can be shared among the CPU and Xeon Phi cores for asynchronous and simultaneous computations on both the CPU and co-processor cores. We split the column computations in such a way that the Xeon and Xeon Phi complete their share of the computations at the same time. The Xeon asynchronously offloads a portion of the columns to the Xeon Phi and proceeds with its own column computations. When the Xeon finishes, results from the Xeon Phi can be collected. This way, a portion of the column computations can be hidden by using asynchronous offloads. The ratio of partitioning is based on the times for computations of the columns on Xeon and Xeon Phi, and further fine-tuned based on experiments with different values of columns/chunk.

Due to our fine-grained offloading of the computations, Xeon-Xeon Phi data transfers are incurred for each offload. We performed a number of optimizations to reduce/hide the data transfer times: \\
{\bf Packing:} The initial version of our proportional partitioning had very huge data transfer times, about 4-5X larger than the computation times on the Xeon Phi. This also resulted in extremely low bandwidth, of about 20 MB/s, on the Xeon-Xeon Phi interconnect while the bus supports up to 8GB/s transfer rates. The primary reason is that our initial version involved multiple transfers of small amount of data (2-5 MB) as a collection of multiple small arrays. Packing these arrays into a single user defined Structure of Arrays greatly increased the achieved bandwidth to about 400-500 MB/s (nearly 20-25x increase) and thus reduced the asynchronous offload set up and wait times. \\
{\bf Offloading only the needed data:} We also avoided resending a large number of scalars that do not change between the offloads. We modified the offload algorithm to send these scalars only in the first timestep, and make them resident for subsequent offloads. \\
{\bf Static in/out variables:} We made the input and output Structure of Arrays as static in order to further reduce the setup and wait overheads. This resulted in at least 10X reduction in data transfer times. \\
The ratio of division of the available columns between the CPUs and co-processors depends on the different number of cores on both the resources, variable performance of the CPU and co-processor, the resolution of the model run and hence the workload of the column computation, and the number of available Xeon Phi cards in a single node. The ratio also depends on the compiler flags used. Use of fast compiler options result in larger performance improvements on Xeon Phi than on Xeon, and hence more columns can be offloaded to Xeon Phi. Use of more Xeon Phi cards can also result in offloading of more columns. We conducted offline experiments to determine these ratios for these different configurations.
These techniques enable our proportional partitioning approach to be scalable to larger number of nodes and cores.

\subsection{Other Optimizations}

\subsubsection{Strength Reduction:} There were certain division operations within a performance critical subroutine of shallow convection routine that calculated some thermodynamic properties. These divisions could be replaced by less-expensive multiplication operations without affecting the result. We applied such strength reduction in various locations of the convection routines. e.g., $a=b/c/c$ was converted to $a=b/(c*c)$.

\subsubsection{Computation and Data Reuse:} The themodynamical properties calculated in the above-mentioned subroutine were found to be in-variant across multiple iterations of the loops in this subroutine. Hence, these calculations can be performed outside the loops and the results can be used for multiple iterations instead of calculating for each iteration, as is done in the original code. Some of these calculations corresponded to costly divisions. By moving these calculations outside the column loop, we observed significant performance improvements.

\subsubsection{Compiler Flags:} Past studies of application optimizations on the Xeon Phi have employed the use of compiler flags to speed up math computations, especially those involving exponent, logarithm and power computations. The Xeon Phi has low-precision hardware support to accelerate these computations and the compiler flags provide means to specify the needed precision. Apart from speeding up these transcendental math functions, there is also support for FMA (Fused Multiply and Add) on the Xeon Phi.
These are hardware instructions that are generated only when using high speed compiler flags and not in the default flags, since they involve a change in the order of floating point calculations being performed and may not produce bit-by-bit identical results.

The shallow and deep convection loops, especially deep convection, are prime candidates for use of high speed compiler flags due to the presence of large number of power and log computations in performance sensitive loops. By using the ``$-fp-model$ $fast=2$'' compiler flags along with varying precisions, we were able to significantly improve performance of both the loops on the Xeon Phi.

\section{Experiments and Results}
\label{exp_res}

\subsection{Experimental Setup}

In our experiments, we used CESM v 1.2.2 with compset FC5 comprising active atmosphere component with CAM5 physics, active land and sea-ice components, stub land-ice, and data model for ocean. We used three different resolutions in our experiments. Table \ref{resolutions} shows the parameters for the resolutions used in the atmosphere model. In all cases, 30 vertical levels were used.  The performance-related experiments were conducted with 5-day simulation runs.

The experiments were conducted on a cluster containing 8 nodes of 16-core (dual octo-core) Intel Xeon E5-2670 CPU with a speed of 2.6 GHz. Each node is equipped with two Intel Xeon Phi 7120 PX card, each with 61 cores.  In all cases, -O3 compiler optimization was used. For performance-related experiments, we used single-node with 16 threads on the CPU for the default R1 resolution due to the small size involved. For the higher R2 and R3 resolutions, we used all the 128 CPU cores of the cluster with 8 MPI tasks and 16 OpenMP threads per MPI task. We used the Xeon Phi cards depending on the experiment. For experiments involving offloading to the Xeon Phi, 180 threads were used for R1 resolution, and 240 threads for R2 and R3 resolutions. All calculations including those on Xeon-Phi were performed using double precision.

Each result shown was obtained as a mean of five runs. The timings were found to be consistent across the runs with the overall CESM times varying between 1 to 5\%. For CESM times, we report the total time excluding the initialization time.

\begin{table}
 \small
 \centering
 \begin{tabular}{||p{0.65in}|p{0.5in}|p{0.9in}|p{0.4in}||}
  \hline\hline
Pseudonym & Resolution & lat x lon (degrees) & Columns \\
  \hline\hline
R1 (default) & f19\_g16 & 1.9 x 2.5 & 13824 \\ \hline
R2 & f05\_g16 & 0.47 x 0.63 & 221184 \\ \hline
R3 & f02\_g16 & 0.23 x 0.31 & 884736 \\
  \hline\hline
 \end{tabular}
\caption{Details of Resolutions}
\label{resolutions}
\end{table}

\subsection{Results on Correctness}

We first demonstrate the correctness of our code modifications due to various optimizations. We verified the accuracy of the results by finding the error growth of the temperature values produced in the code over the simulations \cite{rosinski-roundingerrors-siamjscicomp1997}. The error growth curves compare the RMS difference between the results of the original code and the results of the modified code due to our optimizations, and the RMS difference between the results from the original code and the results obtained by perturbing the inputs by the least significant bit. We refer to these perturbations to the least significant bit as {\em induced perturbations}. In general, for an optimization or modification to be acceptable, the error growth curve due to the modification should be smaller than the error growth curve due to the induced perturbations. Our error growth curves were obtained for a 2-day simulation run with the default R1 resolution.

\begin {figure}
\centering
\includegraphics[scale=0.35]{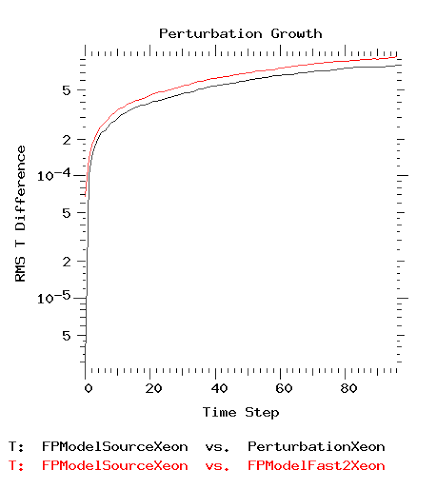}
\caption{Error Growth on Xeon with ``$-fp-model fast=2$'' Compiler Flag}
\label{error_fast2_xeon}
\end{figure}

We first show the error growths with using the advanced compiler flag of ``$-fp-model fast=2$'', which is expected to give fast but less accurate results, over the default flag of ``$-fp-model source$'', which uses source precision. Figure \ref{error_fast2_xeon} shows the error results on Xeon. We find that the use of the ``$-fp-model fast=2$'' optimization did not alter the accuracy significantly. Hence the advanced compiler flags can be safely used to potentially obtain high performance without compromizing on accuracy.

\begin {figure}
\centering
\includegraphics[scale=0.35]{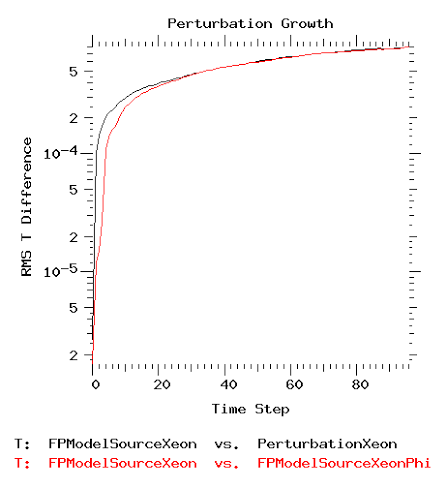}
\caption{Error Growth with Execution on Xeon Phi with ``$-fp-model source$'' Compiler Flag}
\label{error_xeonphi}
\end{figure}

Figure \ref{error_xeonphi} shows the error growth for our modified optimized code executing on Xeon Phi compared with the original code on Xeon. We find that our offloading to Xeon Phi did not alter the error, and the error growth matches well with the induced perturbations.

\subsection{Performance Results}

For most of the performance-related results, we used the medium-range R2 resolution run using 8 MPI tasks on 8 nodes with 16 threads per MPI task for a total of 128 threads. For the Xeon Phi related experiments, we used one Xeon Phi in each node for a total of 8 Xeon Phi cards.

\subsubsection{Performance Benefits of Individual Optimizations}

We first show the performance benefits due to the individual optimizations.

%-----------------------------------------

Table \ref{firstprivate_reduction} shows the timings of the loop before and after reducing first private for scalars on both Intel and Xeon Phi. The experiments correspond to 10-day CESM run with $f16\_g19$ resolution. Using columns 2-4, we find that the use of firstprivate reduction related optimization reduced both the OpenMP overheads for memcpy by about 4.5\% in both Xeon and Xeon Phi, and also reduced the resulting overall looptimes by about 4.8\% in Xeon and about 6.0\% in Xeon Phi. While the OpenMP loop overhead on Xeon considerably reduced to only 2.3\% on Xeon, it is still a significant 20\% in Xeon Phi. The last column of the table shows that the primary reason for this large overhead in Xeon Phi is OpenMP dynamic scheduling. The use of static scheduling reduced the overhead significantly to 7.7\%.
% Thus, improvements in OpenMP runtime for dynamic scheduling will have to be strongly considered in the future versions of Intel Xeon Phi architecture.
While static scheduling reduced the OpenMP loop overhead, we find that the overall loop runtime with dynamic scheduling is about 40\% less than with static scheduling.

\begin{table*}
 \small
 \centering
 \begin{tabular}{||p{1.5in}|p{0.75in}|p{0.75in}|p{0.75in}|p{0.75in}|p{1.5in}||}
\hline\hline
Runtime Component & \multicolumn{2}{|p{1.5in}|}{Intel Xeon with 16 threads, dynamic scheduling} & \multicolumn{2}{|p{1.5in}|}{Intel Xeon Phi with 180 threads, dynamic scheduling} & Intel Xeon Phi with 180 threads, static scheduling \\ \hline
  & Original & After firstprivate reduction & Original & After firstprivate reduction & After firstprivate reduction \\ \hline\hline
Loop runtime (msecs) & 16.26 & 15.47 & 49.16 & 46.20 & 66.88 \\ \hline
Avg. thread runtime (msecs) & 15.15 & 15.11 & 36.9 & 36.7 & 61.70 \\ \hline
\% Overhead & 6.8\% & 2.3\% & 25.0\% & 20.5\% & 7.7\% \\ \hline\hline
  \end{tabular}
\caption{Firstprivate Reduction and Dynamic Scheduling in Shallow Convection}
\label{firstprivate_reduction}
\end{table*}

Our optimization due to elimination of false sharing resulted in 7\% improvement in single-thread performance, 18\% improvement in performance with 16-threads on Xeon, and 30-40\% improvement with 180 threads in Xeon Phi. The code rearrangement for elimination of false sharing also resulted in increase in scalability of the loops from about 37\% to 64-79\% with 16 threads on Xeon.

The combination of strength reduction and data reuse resulted in improvements of 21.5\% on Xeon and 3-10\% on Xeon Phi for single and multi-thread executions.

Tables \ref{all_fpmodel_xeon}-\ref{all_fast2_xeonphi} show and summarize the benefits of all our optimizations for a model run corresponding to $f19\_g16$ resolution, single-node run with 16 OpenMP threads for 16-core Xeon CPU and 240 threads for Xeon Phi, with 13824 columns across the entire grid divided into 16 chunks of 864 columns each, and using -O3 compiler optimization. The results for a row correspond to the cumulative optimizations from the previous rows and the optimization for the current row. The non-convections parts were threaded on the Xeon CPU at the chunk level, and the convection routines threaded at the fine-grained loop level either on the Xeon CPU or on Xeon Phi. The CESM model was executed with a 5-day run for collecting timings and a 2-day run for perturbation. The first two tables show results with compiler flag ``$-fp-model source$'' and the last table shows results with compiler flag ``fast=2''.

\begin{table*}
 \small
 \centering
 \begin{tabular}{||p{1.5in}|p{0.75in}|p{0.75in}|p{0.75in}|p{0.75in}|p{0.75in}||}
\hline\hline
Optimization &	Shallow Convection loop runtime	& Deep Convection loop runtime	& Convection runtime & Atmosphere runtime & CESM runtime \\ \hline\hline
Baseline & 120.6 & 19.1	& 284.3	& 830.1	& 1030.8 \\ \hline
Firstprivate overhead optimization in shallow convection & 78.3	& 19.1 & 238.4 & 786.9 & 984.4 \\ \hline
Dynamic scheduling for load balance & 66.5 & 17.1 & 226.2 & 776.4 & 972.1 \\ \hline
Eliminating false sharing in shallow convection loop & 54.3 & 17.1 & 223.1 & 773.0 & 970.0 \\ \hline
Strength Reduction & 42.6 & 17.1 & 211.6 & 762.8 & 960.0  \\\hline\hline
  \end{tabular}
\caption{All optimizations - ``-fp-model source'' on Xeon. All times are in seconds}
\label{all_fpmodel_xeon}
\end{table*}

\begin{table*}
 \small
 \centering
 \begin{tabular}{||p{1.5in}|p{0.75in}|p{0.75in}|p{0.75in}|p{0.75in}|p{0.75in}||}
\hline\hline
Optimization &	Shallow Convection loop runtime	& Deep Convection loop runtime	& Convection runtime & Atmosphere runtime & CESM runtime \\ \hline\hline
Baseline & 631.2 & 62.1 & 1410.8 & 1936.3 & 2136.9 \\ \hline
Firstprivate overhead optimization in shallow convection & 469.4 & 62.1	& 1256.6 & 1782.8 & 1984.8 \\ \hline
Dynamic scheduling for load balance & 409.8 & 49.0 & 1190.8 & 1718.9 & 1915.7 \\ \hline
Eliminating false sharing in shallow convection loop & 288.3 & 49.0 & 1073.5 & 1615.4 & 1811.1 \\ \hline
Strength Reduction & 258.1 & 49.0 & 1053.4 & 1590.7 & 1787.1  \\ \hline\hline
  \end{tabular}
\caption{All optimizations - ``-fp-model source'' on Xeon Phi. All times are in seconds}
\label{all_fpmodel_xeonphi}
\end{table*}

\begin{table*}
 \small
 \centering
 \begin{tabular}{||p{1.5in}|p{0.75in}|p{0.75in}|p{0.75in}|p{0.75in}|p{0.75in}||}
\hline\hline
Optimization &	Shallow Convection loop runtime	& Deep Convection loop runtime	& Convection runtime & Atmosphere runtime & CESM runtime \\ \hline\hline
Baseline & 520.0 &43.7 & 1291.5 & 1774.9 & 1973.4 \\ \hline
Firstprivate overhead optimization in shallow convection & 363.8 & 43.1 & 1195.2 & 1828.6 & 2094.1 \\ \hline
Dynamic scheduling for load balance & 325.7 & 35.5 & 1095.0 & 1581.7 & 1775.1 \\ \hline
Eliminating false sharing in shallow convection loop & 198.6 & 35.5 & 961.5 & 1453.3 & 1649.4 \\ \hline
Strength Reduction & 192.3 & 35.5 & 955.3 & 1446.0 & 1640.0 \\ \hline\hline
  \end{tabular}
\caption{All optimizations - ``-fp-model fast=2 -fimf-precision=high'' on Xeon Phi. All times are in seconds}
\label{all_fast2_xeonphi}
\end{table*}

Comparing the first and last lines in Tables \ref{all_fpmodel_xeon} and \ref{all_fpmodel_xeonphi}, we find that on Xeon CPU with ``$-fp-model source$'' our optimizations result in performance improvements of about 65\% for shallow convection loop, 10\% for deep convection loop, 25\% for the entire convection, 8\% for the entire atmosphere model, and 7\% improvement in the overall CESM model. On Xeon Phi with ``$-fp-model source$'' our optimizations result in performance improvements of about 59\% for shallow convection loop, 21\% for deep convection loop, 25\% for the entire convection, 18\% for the entire atmosphere model, and 16\% improvement in the overall CESM model. Thus, we find that our optimization provide higher returns for the atmosphere model and the entire CESM run on the Xeon Phi co-processor. Comparing the last lines of the Tables \ref{all_fpmodel_xeonphi} and \ref{all_fast2_xeonphi}, we find that the use of the fast compiler flags provided further performance improvements of about 25\% for shallow convection, 28\% for deep convection, 9\% for total convection, 9\% for the entire atmosphere model, and 8\% for the overall CESM run.

%---------------------------------------

\begin {figure}
\centering
\includegraphics[scale=0.35]{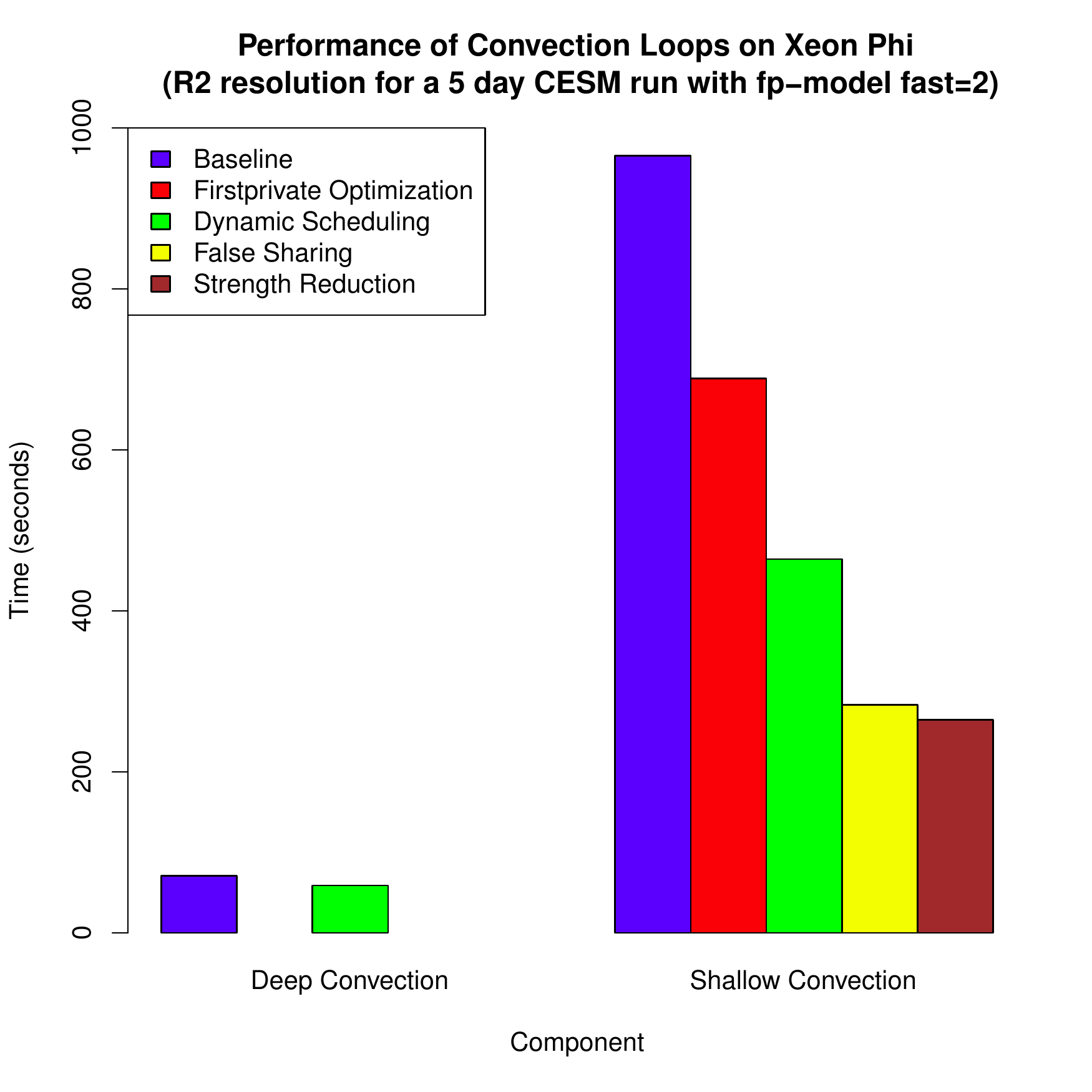}
\caption{Effect of Optimizations on Deep and Shallow Convection Execution Times on Xeon Phi for ``fast=2'' flag for R2 resolution}
\label{opt_deepshallow_perf_xeonphi}
\end{figure}

\begin {figure}
\centering
\includegraphics[scale=0.35]{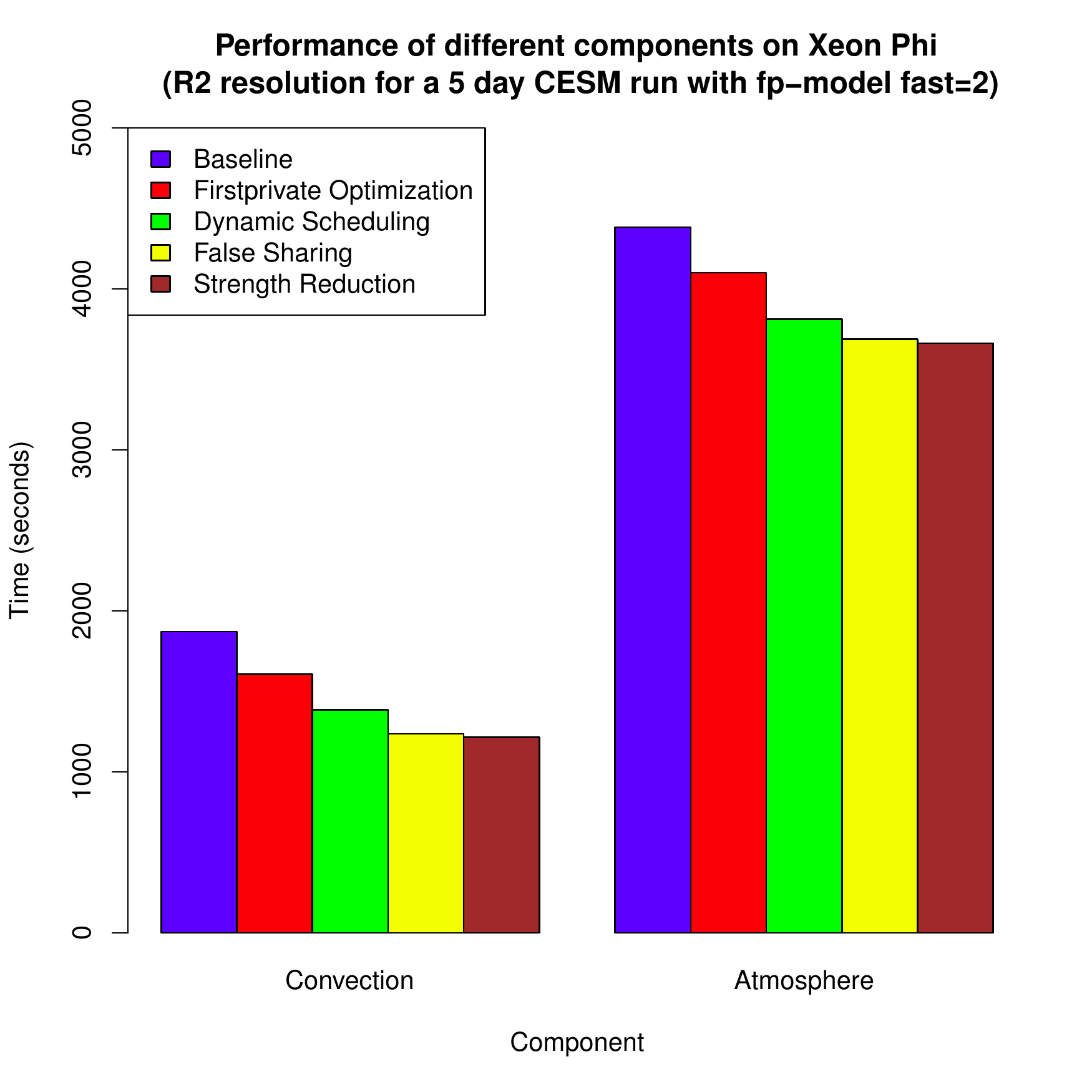}
\caption{Effect of Optimizations on Convection, Atmosphere Execution Times on Xeon Phi for ``fast=2'' flag for R2 resolution}
\label{opt_conatmcesm_perf_xeonphi}
\end{figure}

We first show the performance benefits due to the individual optimizations.
For the results in Figures \ref{opt_deepshallow_perf_xeonphi} and \ref{opt_conatmcesm_perf_xeonphi}, the optimizations are cumulatively applied in the order shown.

\begin {figure}
\centering
\includegraphics[scale=0.35]{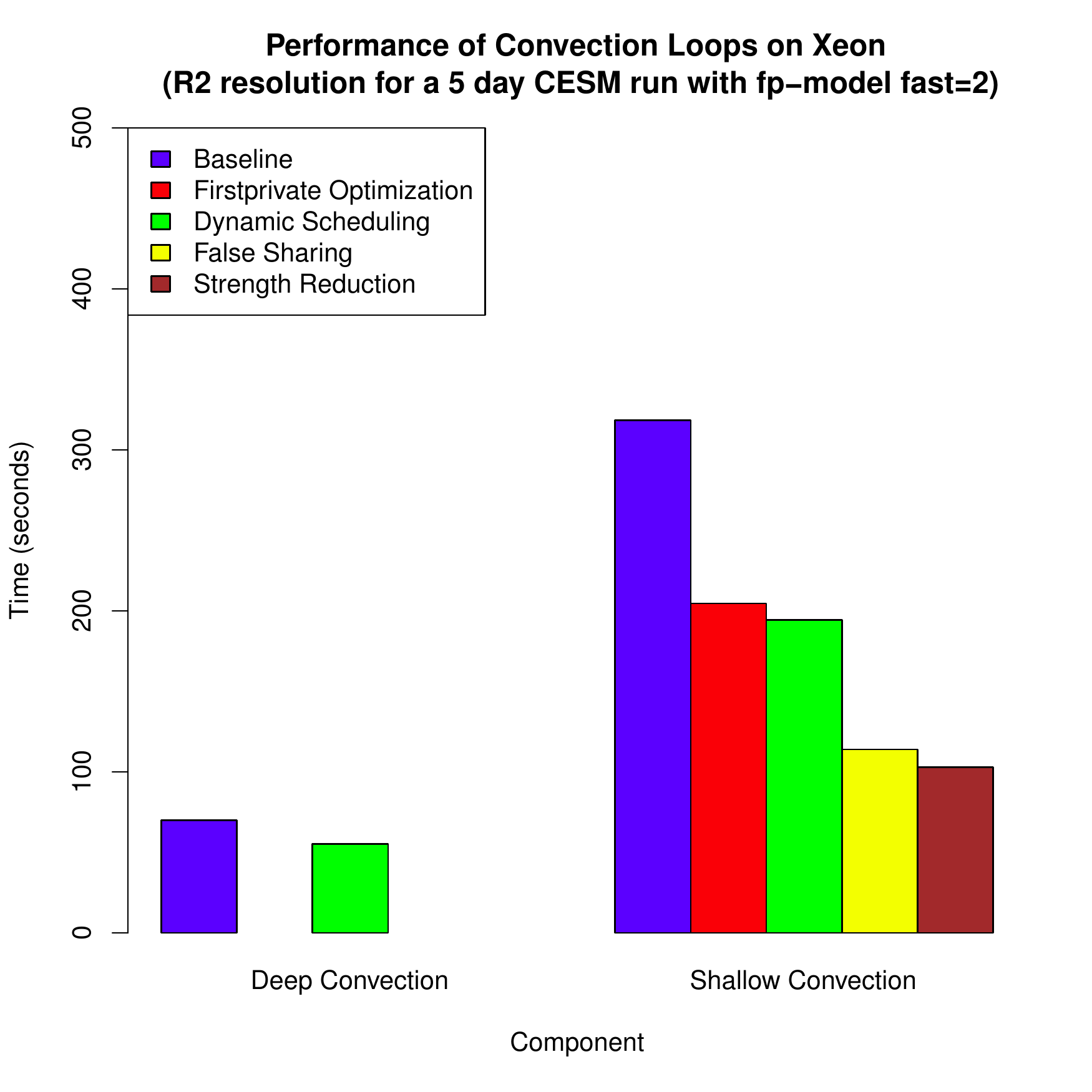}
\caption{Effect of Optimizations on Deep and Shallow Convection Execution Times on Xeon for ``fast=2'' flag for R2 resolution}
\label{opt_deepshallow_perf_xeon}
\end{figure}

 Figure \ref{opt_deepshallow_perf_xeon} shows the improvements in shallow and deep convections times over the baseline model on Xeon for the different optimizations for the R2 resolution. Note that only the dynamic scheduling optimization is applicable to deep convection, and it results in 20\% performance improvement for this convection routine. For shallow convection, the performance improvements are 34\% with firstprivate reduction, an additional 5\% with dynamic scheduling, an additional 25\% with false sharing elimination, and an additional 5\% with strength reduction. Thus we find firstprivate reduction and false sharing elimination as the most important optimizations in our work. Note that these optimizations relate to data management among multiple threads.

Figure \ref{opt_deepshallow_perf_xeonphi} shows the improvements in shallow and deep convections times over the baseline model on Xeon Phi for the different optimizations for the R2 resolution. Note that only the dynamic scheduling optimization is applicable to deep convection, and it results in 17\% performance improvement for this convection routine.
 Dynamic scheduling results in 17\% improvement in the deep convection.
For shallow convection, the performance improvements are 30\% with firstprivate reduction, an additional 24\% with dynamic scheduling, an additional 16\% with false sharing elimination, and an additional 2\% with strength reduction.
Interestingly, we find that unlike in Xeon,
dynamic scheduling plays a major role in addition to the firstprivate reduction and false sharing elimination in Xeon Phi. Thus, load balancing using dynamic scheduling is important in Xeon Phi due to the use of a large number of threads. Note that the firstprivate reduction and false sharing elimination relate to data management among multiple threads.

\begin {figure}
\centering
\includegraphics[scale=0.35]{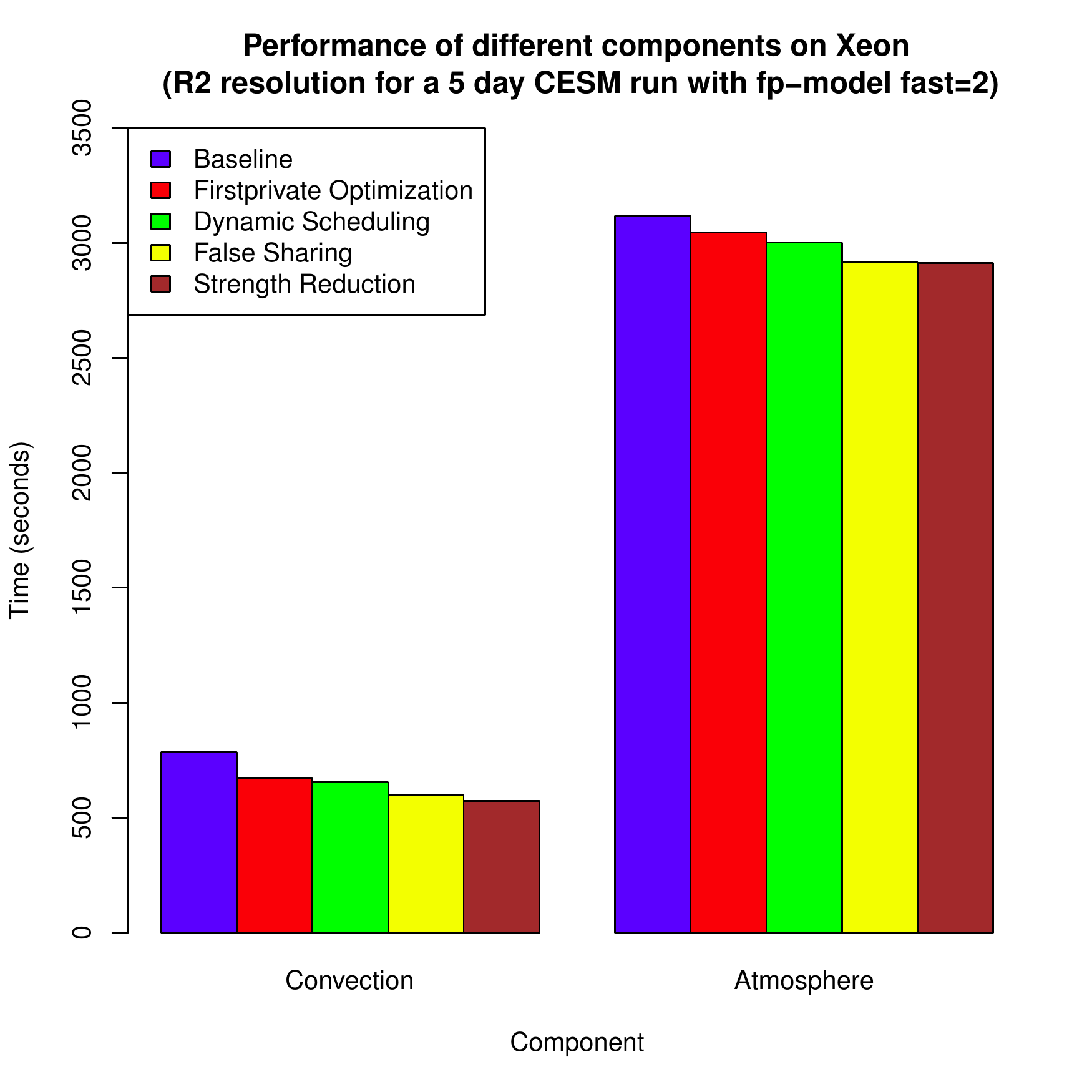}
\caption{Effect of Optimizations on Convection, Atmosphere Execution Times on Xeon for ``fast=2'' flag for R2 resolution}
\label{opt_conatmcesm_perf_xeon}
\end{figure}

We next show the effects of the individual optimizations on the overall convection and atmosphere execution times.  Figures \ref{opt_conatmcesm_perf_xeon} and \ref{opt_conatmcesm_perf_xeonphi} show the improvements in the total convection and atmosphere times over the baseline model on Xeon and Xeon Phi, respectively, for the different optimizations for the R2 resolution. We notice that the individual optimizations in the deep and shallow convection show significant and visible individual improvements even in the higher-level computations in the call trace, namely, overall convections and complete atmosphere model executions.  For example in Xeon Phi, the firstprivate reduction and dynamic scheduling optimizations result in performance improvements of 7\% each in the atmosphere model timings.

Our proportional partitioning method takes into account the variable performance ratios of Xeon and Xeon Phi to offload the column computations of shallow convections to Xeon Phi. Table \ref{pp_results} compares the time taken using proportional partitioning with the times taken for Xeon-only and Xeon-Phi only computations for the R2 and R3 resolutions. The results correspond to 2-day simulation runs. The last two columns of the table also show the times spent on Xeon and Xeon Phi in the proportional partitioning approach. The difference between the total time (4th column) and the maximum of the last two columns gives the overheads including data transfers.

\begin{table}
\centering
\begin{tabular}{|p{0.4in}|p{0.43in}|p{0.65in}|p{0.43in}|p{0.45in}|p{0.6in}|}
\hline\hline
{\bf Resolution} & {\bf Xeon-only} & {\bf Xeon-Phi Only} & \multicolumn{3}{|c|}{{\bf Proportional Partitioning}} \\
\hline
 & & & Total Time & Xeon Time & Xeon Phi Time \\
\hline\hline
R2 & 38.0 & 83.6 & 88.0 & 46.4 & 27.2 \\
\hline
R3 & 292.5 & 744.3 & 421.6 & 262.7 & 164.0 \\
\hline\hline
\end{tabular}
\caption{Results of Proportional Partitioning. All times are in seconds}
\label{pp_results}
\end{table}

First, we find that the Xeon Phi shows 2X slowdown when compared to the Xeon CPU. This is primarily due to the poor single thread performance on Xeon Phi, and the lack of vectorization opportunities in the critical loops of the shallow and deep convection routines, in which convergence is tested for termination. We expect the performance to improve in the next generation Intel Xeon Phi Knoghts Landing processors that have superior single thread performance.

As for the proportional partitioning approach, it gives equivalent (for R2) or 1.77x performance improvement over the Xeon-Phi only computations for both the resolutions. This is due to some of the work that is shared by the Xeon CPU. When compared to the Xeon-only approach, the proportional partitioning shows slowdowns primarily due to the data transfer overheads for the fine grained offloads on the PCIe link between the host and the coprocessor. Our study involves fine-grained parallelism in Xeon Phi in which the computations are offloaded at the column level and not at the chunk level. For every chunk assigned to a node, offloading is performed once. In all our experiments, 16 chunks are assigned to a Xeon node to optimize the computations of the other physics routines. This results in 16 offloads per time step and the corresponding large data transfer overheads. In future, we plan to explore chunk-level offloading in which only one offloading will be performed for all the chunks assigned to a node. Interestingly, we find that this slowdown in proportional partitioning when compared to the Xeon-only result decreases as we increase the resolution or problem size: 132\% for R2 and 44\% for R3. Thus, the proportional partitioning approach has very good promise as the climate modeling community plans to explore large and very large resolutions in the future.

\begin {figure}
\centering
\includegraphics[scale=0.35]{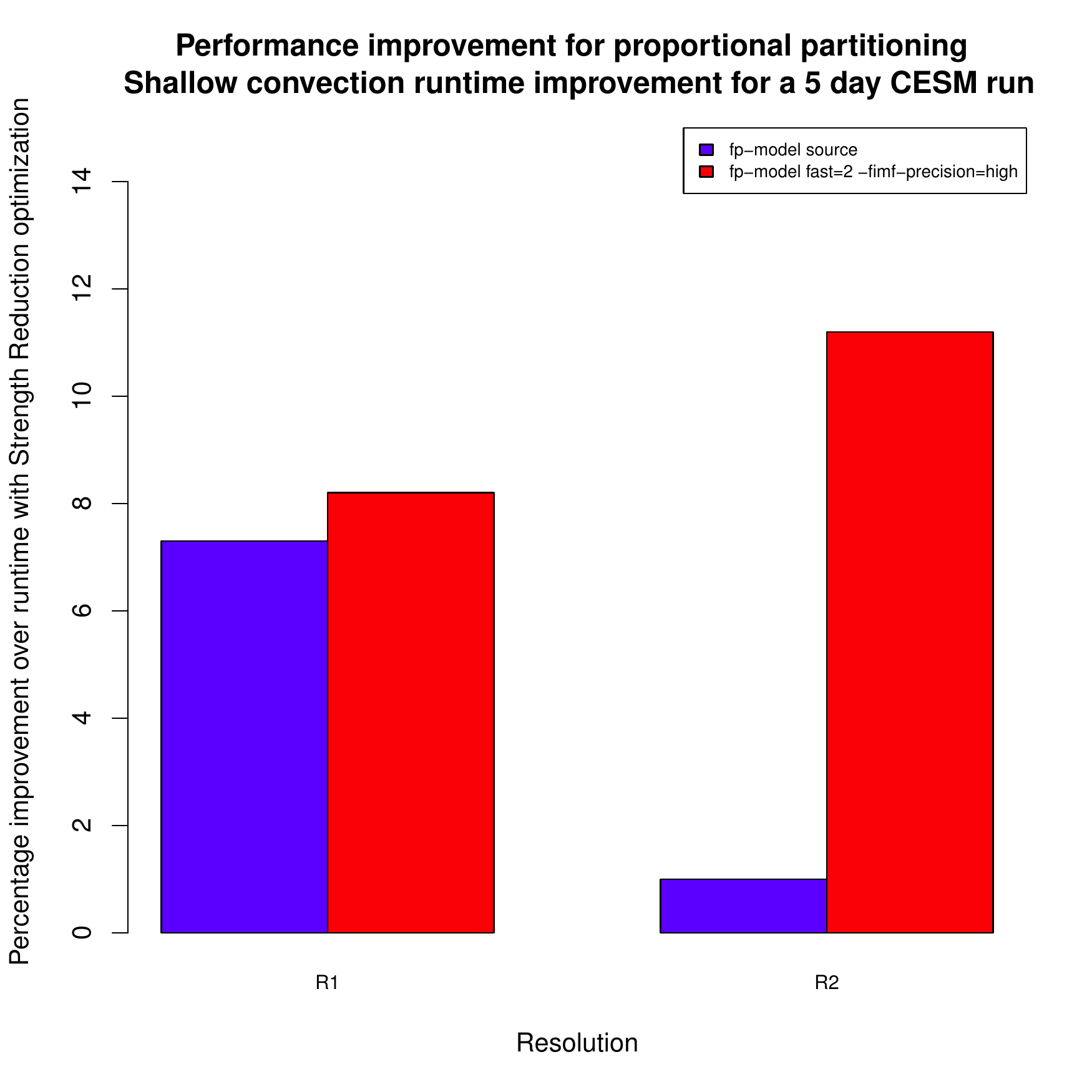}
\caption{Benefits of Proportional Partitioning with ``$-fp-model source$'' and ``fast=2'' flags and for R1 and R2 resolutions}
\label{proportional_part_perf}
\end{figure}

\subsubsection{Overall Optimization Benefits}

\begin {figure}
\centering
\includegraphics[scale=0.35]{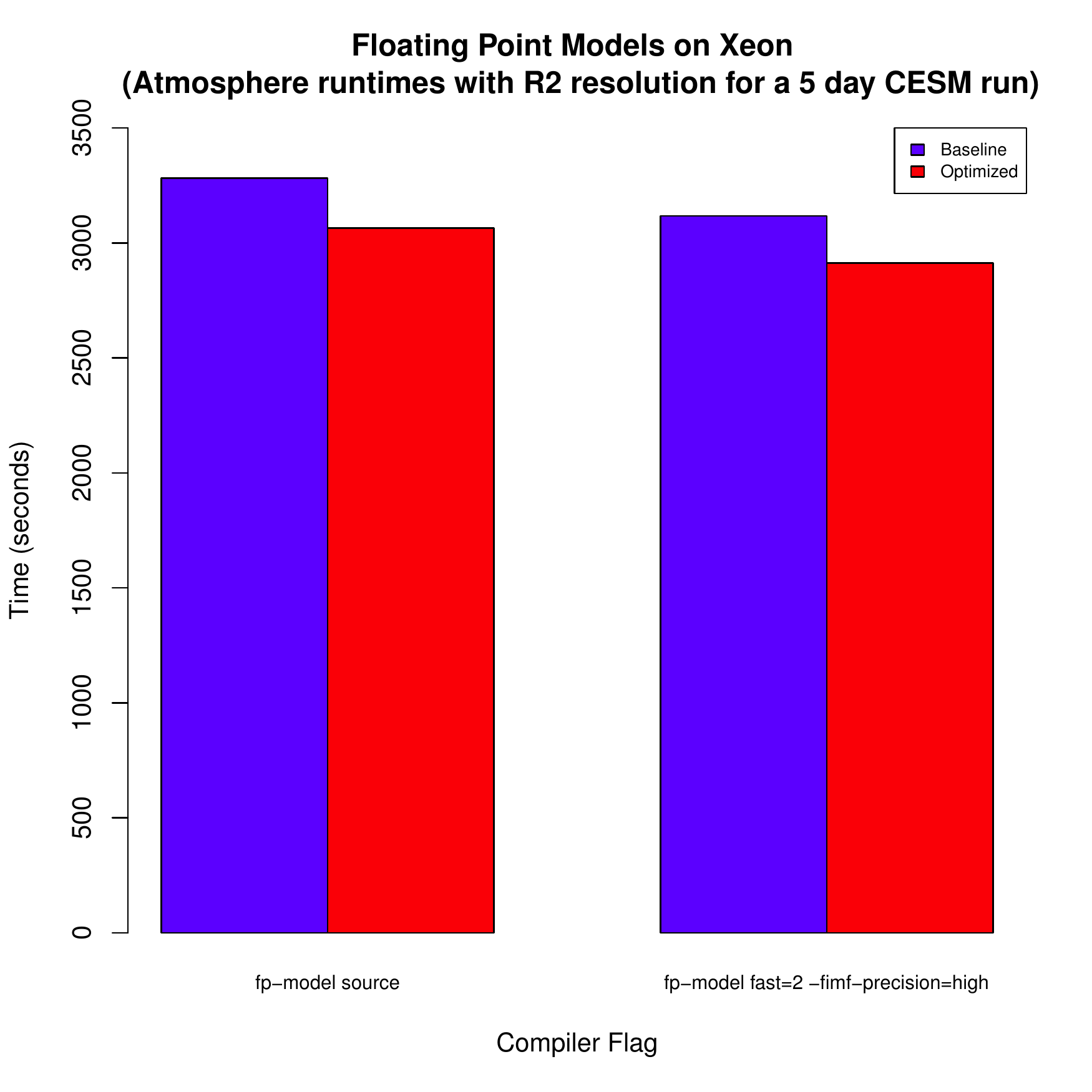}
\caption{Execution Times for Atmosphere on Xeon for different compiler flags for R2 resolution}
\label{flags_xeon_perf}
\end{figure}

Figure \ref{flags_xeon_perf} shows the performance improvement in the complete atmosphere calculations of the modified code containing our optimizations over the baseline code on Xeon CPU for the two compiler flags namely, ``$-fp-model source$'' and ``$-fp-model fast=2$'' flags. The modified code contains all optimizations except proportional partitioning which involves Xeon Phi. We find that the performance improvement is about 5\% with both the compiler flags.

\begin {figure}
\centering
\includegraphics[scale=0.35]{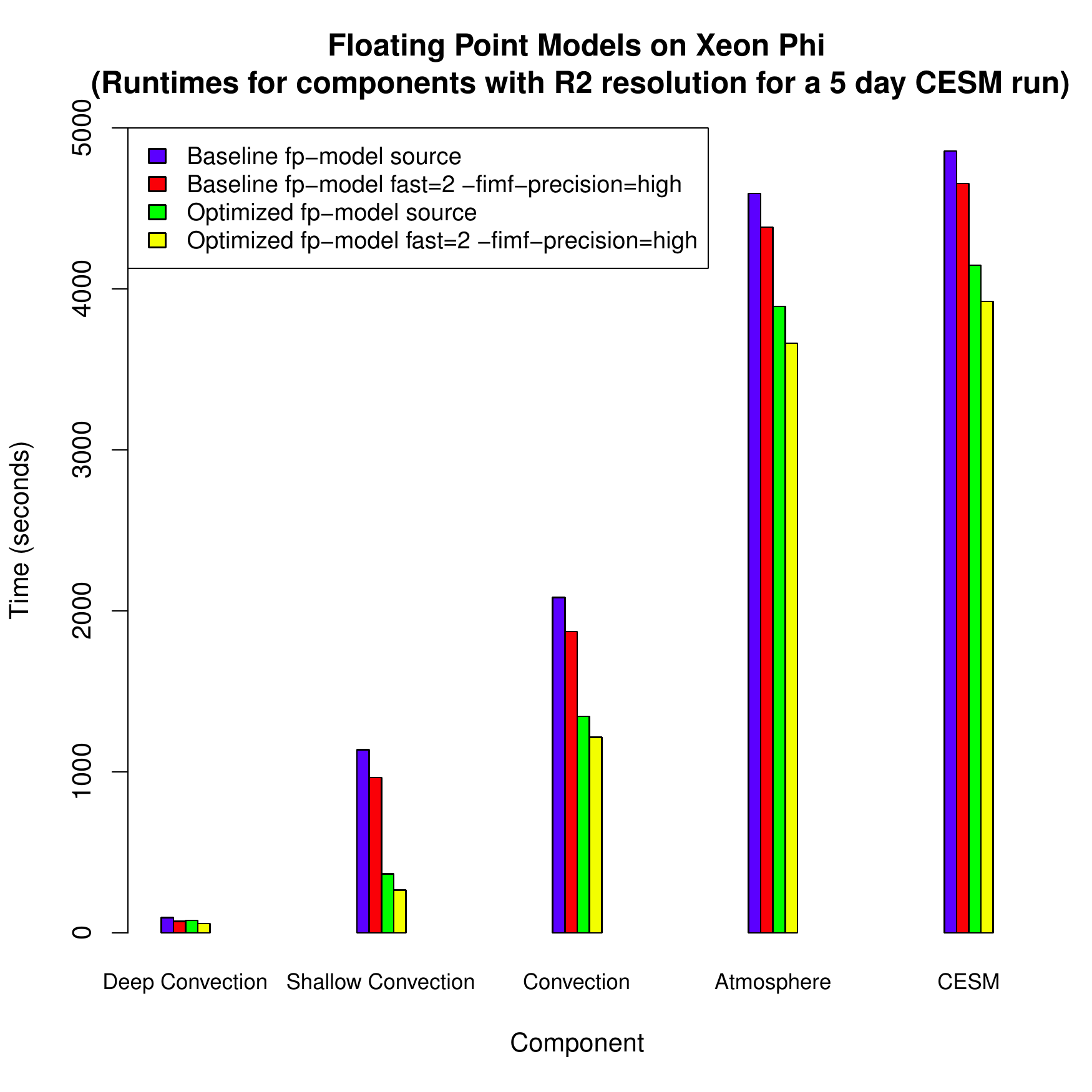}
\caption{Execution Times on Xeon Phi for different compiler flags for R2 resolution}
\label{flags_xeonphi_perf}
\end{figure}

We show the effect of all the optimizations put together for the various modules. We first show the impact of compiler flags on our optimizations.
Figure \ref{flags_xeonphi_perf} shows the performance improvements in the different components due to our optimizations over the baseline code on Xeon Phi for the ``$-fp-model source$'' and ``$-fp-model fast=2$'' flags. We find that the performance improvements with the ``source'' flag are 19\%, 2.5x, 30\%, 15\%, 16\% for the deep convection, shallow convection, total convection, total atmosphere and the entire CESM, respectively. With the ``fast=2'' flag, the performance improvements are  17\%, 4x, 37\%, 17\%, 16\% for the deep convection, shallow convection, total convection, total atmosphere and the entire CESM, respectively. We find that the use of ``fast=2'' flag results in higher performance benefits due to our optimizations than the use of ``$-fp-model source$'' flag in the shallow convection calculations (2.5x vs 4x). This is due to our larger number of vectorization-related optimizations in the shallow convection routines, whose benefits are higher with the fast flags. The use of ``fast=2'' flags generates hardware instructions for transcendental math functions on Xeon Phi which improves performance.
Comparing this with the previous figure, we also find that
 the performance benefits with our optimizations are much higher in Xeon Phi than in Xeon. This is due to the wider vectorization units in the Xeon that can harness our vectorization related optimizations more. Significantly, we find that our work on optimizations in the convection routine gives about 15\% improvement even in the overall CESM executions.

\begin {figure}
\centering
\includegraphics[scale=0.35]{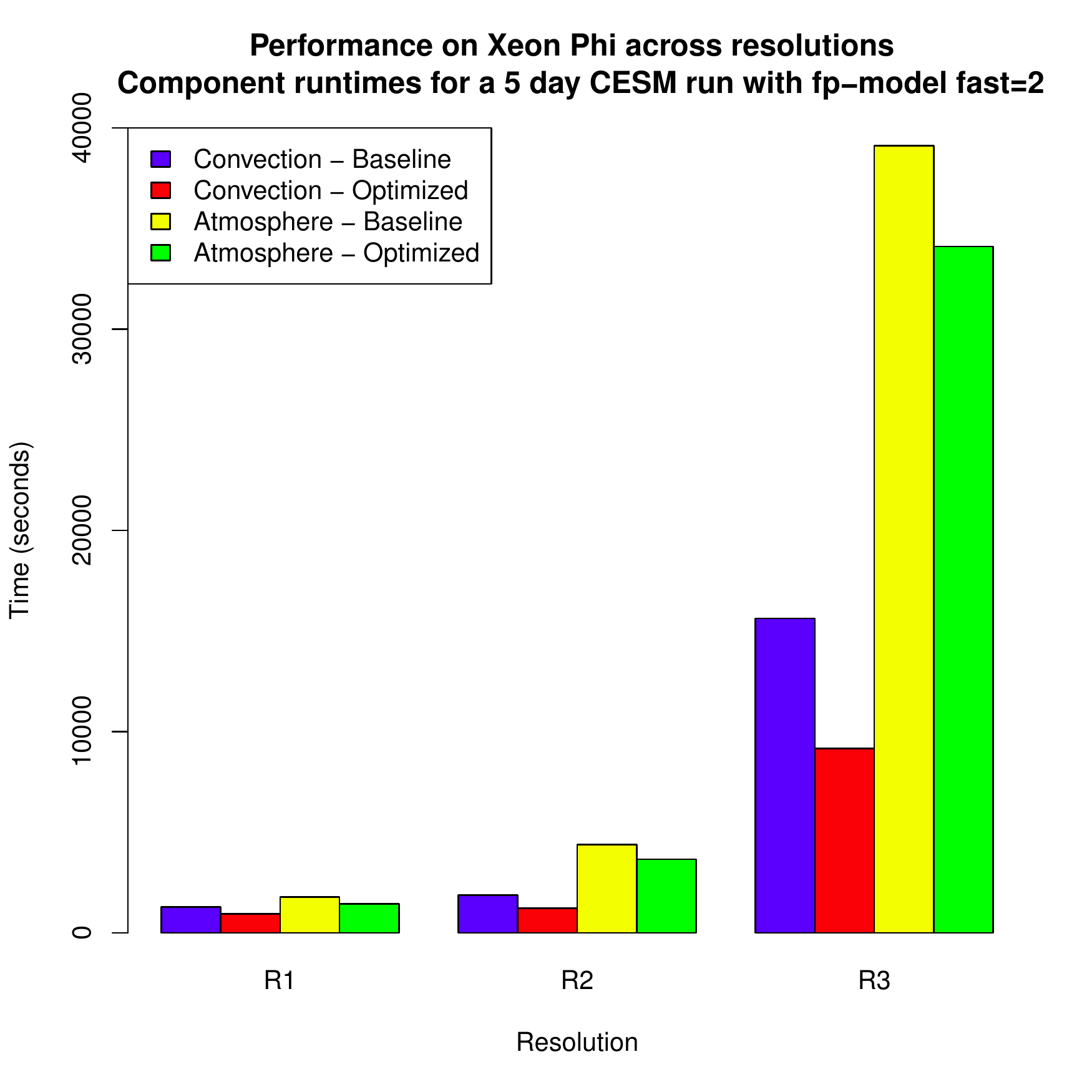}
\caption{Execution Times for Different Resolutions on Xeon Phi for ``fast=2'' flag. R1 was executed on a single node, 16 cores. R2 and R3 were executed on 8 nodes, 128 cores.}
\label{res_perf}
\end{figure}

We also evaluate the benefits of all our optimizations for different resolutions. Figure \ref{res_perf} compares the execution times of the entire convection and atmosphere calculations with the baseline model on Xeon Phi for the three resolutions, R1, R2 and R3. We find that as the resolution is increased, the performance gains due to our optimizations increase for convection calculations. Specifically, the performance gains in the convection calculations are 26\%, 35\% and 42\% for R1, R2, and R3 resolutions, respectively. Thus, overall, we find that optimizations will play bigger roles in the future when the climate models will explore larger resolutions and many-core systems will be built with wider vectorization units with more advanced compiler flags.

\subsection{Savings for Multi-Century Runs}

Multi-century simulation runs are typically of interest in climate models to study the long-term effects on the climate due to various factors including CO$_{2}$ levels. We extrapolated the performance gains obtained due to our optimized convection computations for a multi-century simulation run using our runs for limited number of simulation days. Specifically, we obtained the performance gains in seconds on both Xeon and Xeon Phi over their baseline executions for a 5-day simulation run, and extrapolated the gains in terms of days for a 1000-year simulation run. Table \ref{multi_century_savings} shows savings in terms of the number of days for execution for different resolutions.

\begin{table}
\small
\centering
\begin{tabular}{|p{0.8in}|p{1.0in}|p{1.0in}|p{1.0in}|}
\hline
Resolution & Savings on Xeon (days) & Savings on Xeon Phi (days) \\
\hline
R1 & 61.4 & 284.0 \\
R2 & 181.0 & 555.10 \\
R3 &  1349.70 & 5467.40 \\
\hline
\end{tabular}
\caption{Savings in Execution Days for Multi-Century Runs with``$-fp model fast=2 -fimf-precision=high$'' Compiler Flag}
\label{multi_century_savings}
\end{table}

We find that the use of our optimizations results in highly significant savings in execution days.
As shown above, the performance and hence the savings increase with increase in resolutions. We also find that the savings on Xeon Phi are about 4X the savings on Xeon. This is due to our optimizations harnessing the wider vector units of the Xeon Phi.
Considering one of the results, for example on Xeon, our optimized executions results in savings of up to 181 days or half a year in execution for the R2 resolution. These are highly significant savings and implies not only improved performance, but also savings in power consumption, electricity and maintenance costs.

\section{Discussion}
\label{discussion}

While we had primarily worked on and demonstrated our results for a climate modeling application, we can derive some general principles that can be applied for other scientific applications.

The shallow convection loop that we consider in our study of fine-grained parallelism on the Xeon Phi, displayed loop level data parallelism involving hundreds of multi-dimensional variables. Our optimization related to data management served to illustrate the importance of carefully choosing and designing the OpenMP clause (dynamic vs static scheduling, data management clauses), not only for correctness but also for performance. While studying the scalability of the unoptimized version, we noted that designating variables as firstprivate led to poor scalability of the loop on the Xeon, and this negative effect was vastly amplified on the Xeon Phi. The underlying cause is the bottleneck created by the memory copies that are used to implement this clause. These challenges will be widely seen in future large-scale scientific applications when deployed on the widely prevalent large-scale heterogeneous architecture. A generic tool that automatically chooses the OpenMP clause based on performance will be highly useful for the scientific community. Building such a tool can be part of our future work.

Based on the results in Figures \ref{opt_deepshallow_perf_xeon} and \ref{opt_deepshallow_perf_xeonphi} and Table \ref{pp_results}, we find that the Xeon Phi shows 2X slowdown when compared to the Xeon CPU. This is primarily due to the poor single thread performance on Xeon Phi, and the lack of vectorization opportunities in the critical loops of the shallow and deep convection routines, in which convergence is tested for termination. We expect the performance to improve in the next generation Intel Xeon Phi Knoghts Landing processors that have superior single thread performance.

Proportional partitioning of computations to Xeon and Xeon Phi can generally be applied to applications and loops where the co-processor can serve as yet another in-node compute device that can lead to overall reduction in application runtime. Applications that consider fine-grained offloading of parallel computations involving multiple transfers of large number of small arrays should consider packing of the data into structure of arrays.

The Xeon Phi supports hardware instructions for certain transcendental operations like power, exp, etc. It is likely that many scientific codes make heavy use of these operations, but rarely experiment by changing the floating point model used because of the potential impact that it could have on correctness of the simulation output. We performed a study to illustrate the potential performance benefits of using these hardware instructions on the Xeon Phi, and demonstrated about 8\% improvement in performance of shallow convection loop by using a fast-math floating point model, with an acceptable level of accuracy.

\section{Conclusions and Future Work}
\label{con_fut}

In this work, we successfully offloaded the time-consuming deep and shallow convection calculations in the atmosphere model of CESM to Xeon Phi. We performed a number of optimizations including load balancing column computations corresponding to varying cloud cover, reduction of firstprivate variables to reduce data copying overheads, avoiding false sharing, proportional partitioning across CPU and co-processor cores, and strength reduction techniques. Our combined set of optimizations yielded 30\% performance improvements with Xeon Phi. In future, we plan to adopt similar strategies for other components of CESM and provide large-scale improvements for the entire CESM on Intel Xeon Phi architectures.
We also plan to explore our optimizations in the future Intel Xeon Phi architecture of Knights Landing.

\section*{Acknowledgments}
This project is supported by the Intel{\textregistered} Parallel Computing Centre for Modelling Monsoons and Tropical Climate (IPCC-MMTC), India sponsored by the Intel{\textregistered} Corporation.

\bibliographystyle{IEEEtran}
\bibliography{convectionsphi}

\end{document}